**Naval Research Laboratory**
Washington, DC 20375-5320

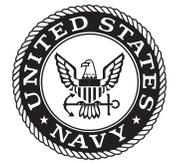

NRL/MR/7165--20-10,195

# Hydraulically Amplified Self-Healing ELectrostatic (HASEL) Inspired Actuators

James P. Wissman*
Alec K. Ikei
Kaushik Sampath
Charles A. Rohde

*Acoustics Signal Processing and Systems Branch*
*Acoustics Division*

*Former National Research Council Postdoc

November 2, 2020

# REPORT DOCUMENTATION PAGE

Form Approved
OMB No. 0704-0188

| 1. REPORT DATE (DD-MM-YYYY) 02-11-2020 | 2. REPORT TYPE NRL Memorandum Report | 3. DATES COVERED (From - To) 09/25/2019 – 09/25/2020 |
|---|---|---|
| 4. TITLE AND SUBTITLE HASEL Inspired Actuators | | 5a. CONTRACT NUMBER |
| | | 5b. GRANT NUMBER |
| | | 5c. PROGRAM ELEMENT NUMBER |
| 6. AUTHOR(S) James P. Wissman, Alec K. Ikei, Kaushik Sampath, and Charles A. Rohde | | 5d. PROJECT NUMBER |
| | | 5e. TASK NUMBER |
| | | 5f. WORK UNIT NUMBER |
| 7. PERFORMING ORGANIZATION NAME(S) AND ADDRESS(ES) Naval Research Laboratory 4555 Overlook Avenue, SW Washington, DC 20375-5320 | | 8. PERFORMING ORGANIZATION REPORT NUMBER NRL/MR/7165--20-10,195 |
| 9. SPONSORING / MONITORING AGENCY NAME(S) AND ADDRESS(ES) National Research Council 500 Fifth Street Washington, DC 20001 | | 10. SPONSOR / MONITOR'S ACRONYM(S) NRC |
| | | 11. SPONSOR / MONITOR'S REPORT NUMBER(S) |

**12. DISTRIBUTION / AVAILABILITY STATEMENT**

**DISTRIBUTION STATEMENT A:** Approved for public release; distribution is unlimited.

**13. SUPPLEMENTARY NOTES**

**14. ABSTRACT**

This report presents research conducted on amplified self-healing electrostatic (HASEL) actuators. HASEL actuators are comprised of a dielectric fluid sealed between two inextensible layers with bonded, flexible electrodes on its outer surface. When charge is applied to the electrodes, Coulomb force compresses the fluid and causes the actuator to contract. In this work a faster, more customizable and convenient way of creating a HASEL actuator is presented, using a laser engraver to heat-seal and cut polypropylene sheets. Using this technique, a hydraulically actuated soft lens is fabricated and demonstrated.

**15. SUBJECT TERMS**

HASEL actuators  Laser engraver
Soft robotics  Lenses

| 16. SECURITY CLASSIFICATION OF: | | | 17. LIMITATION OF ABSTRACT | 18. NUMBER OF PAGES | 19a. NAME OF RESPONSIBLE PERSON Alec K. Ikei |
|---|---|---|---|---|---|
| a. REPORT Unclassified Unlimited | b. ABSTRACT Unclassified Unlimited | c. THIS PAGE Unclassified Unlimited | Unclassified Unlimited | 12 | 19b. TELEPHONE NUMBER (include area code) (202) 404-4816 |

Standard Form 298 (Rev. 8-98)
Prescribed by ANSI Std. Z39.18

This page intentionally left blank.

# CONTENTS

This page intentionally left blank.

# EXECUTIVE SUMMARY

This report presents research conducted on Hydraulically Amplified Self-healing ELectrostatic (HASEL) actuators. HASEL actuators are comprised of a dielectric fluid sealed between two inextensible layers with bonded, flexible electrodes on its outer surface. When charge is applied to the electrodes, Coulomb force compresses the fluid and causes the actuator to contract. In this work a faster, more customizable and convenient way of creating a HASEL actuator is presented, using a laser engraver to heat-seal and cut polypropylene sheets. Using this technique, a hydraulically actuated soft lens is fabricated and demonstrated.

This page intentionally left blank.

# Hydraulically Amplified Self-Healing ELectrostatic (HASEL) Inspired Actuators

## INTRODUCTION

### 1.1 HASEL Actuators

The appeal of soft robotics comes from the need for adaptability and its ability to interface with delicate objects. Applications can be seen in industrial automation or in biomimetics, where robotic grippers can be used to conform to and hold objects that a rigid system would have a difficult time doing. In previous work by Acome et al. [1], the hydraulically amplified self-healing electrostatic (HASEL) actuator is introduced, and its versatility for manipulating delicate objects is shown by holding and lifting an egg. By stacking or linking these actuators, higher levels of deflection and better form-fitting can be obtained, while speed can be increased by amplifying the applied voltage.

HASEL actuators use two flexible layers (called the "shell") to seal dielectric fluid in a fixed volume. By applying opposite charge on the outer surface of each layer, the electric field applies electrostatic force to squeeze fluid from the away from the electrode region. This transition is drawn in black in Fig. 1(a) – (b). This causes pressure to rise in the fluid, changing the shape of the shell from the extended shape into a cylindrical one (Fig. 1(b) – (c)). The change in geometry is based on the shape and compliance of the electrode and enclosing material. The shape of the as-drawn example causes a length contraction to take place in Fig. 1(c) [5]. In contrast with dielectric elastomer actuators (DEA), HASEL actuators rely on the dielectric fluid to separate the charged layers. The use of a fluid increases their tolerance for manufacturing error as well as the upper limit for voltage induced breakdown. Additionally, the use of a dielectric fluid allows for self-healing if breakdown occurs through the fluid, whereas breakdown in a solid elastomer DEA can cause the actuator to fail.

HASEL actuator shells are usually made though hot pressing thermoplastic layers together [5]. This technique uses a metal press with a pre-formed seal design, which means that reconfiguration requires physical retooling, for any change of the design. To improve ease of reconfiguration, the shell of the HASEL actuator has also been sealed using 3D printer hot ends [7]. The technique prescribed in this work attempts to further automate the process and improve design configurability. By using a laser engraver (PLS6MW, ULS) instead of a hot press, speed, design space and convenience are enhanced.

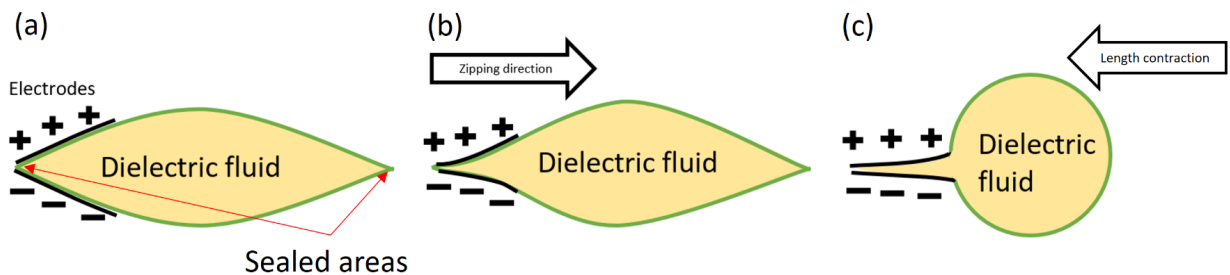

Fig. 1 — Side profile schematic of a HASEL actuator. *a. Electric field is applied to shell. b. Zipping mechanism brings layers of shell together and squeezes fluid away from electrodes. c. Actuator contracts lengthwise, increases height. Figure inspired from reference [5].*

### 1.2   HASEL Lens

Building deformable focusing elements is a popular application of soft robotic technology. Soft lenses have been actuated by non-hydraulic mechanisms, such as through the use of liquid crystal elastomers (LCE) [6], or through electrical activation of an indium tin oxide (ITO) substrate on a polymer film [2]. The main appeal of its softness is that it allows it to change its focal length. The compliance of the lens also has potential to aid in biomedical applications, where form needs to adapt to body shape without damaging its surroundings.

However, hydraulic actuation remains one of the more popular approaches to soft lens actuation. This can be done by applying pressure to a fluid chamber to flex an optical element. For example, Wei, et al. uses a DEA mounted on a frame to apply hydrostatic pressure to change the curvature of a liquid lens, while Xu, et al. use a photopolymer to apply pressure. Although the lens section in both of those works are made from compliant materials, the fluid chamber is not. In comparison, our work uses a completely flexible structure, improving its ability to become a part of a larger soft robotic system.

A recent study by Cheng, et. al showed a variable focus electro-hydraulic lens, which uses encapsulated dielectric fluid between layers of PDMS to provide its focal length change. To make the lens, three circular films of PDMS were cut out, removing a circular section from the central layer for the dielectric fluid. These layers were then adhered together manually using uncured PDMS as an adhesive. Annular, inextensible conductive tape electrodes were attached on the top and bottom PDMS layers. When voltage was applied, the fluid hydraulically deflected the PDMS within the inner diameter of the tape electrodes, which gave a change in focal length. In contrast, the lens fabrication in this work combines the cutting and sealing operations in an automated process using a laser engraver. Additionally, the pumping and lens sections were shown to be separable. Having a separate reservoir allows for more fluid in space limited applications, which may provide larger levels of lens magnification.

### APPROACH

### 1.3   Single Element HASEL Actuator Fabrication

A HASEL actuator has an exterior shell, an interior dielectric fluid, and electrodes on the top and bottom surfaces of the shell. Polypropylene (PP) sheets were used to create the shell, which were taken from an 80um thick, heat-sealable food package (RXF). Leitsilber Conductive Silver Cement (Ted Pella) was used to paint electrodes on the surfaces of the shell, and consumer cooking oil (soybean oil, Market Pantry) was used as the dielectric fluid. A 20kV DC power supply (Stanford Systems, PS375) was used to supply the necessary charge. The sealed area that borders the fluid-filled chamber was necessary to prevent arcing through the air between the top and bottom electrodes.

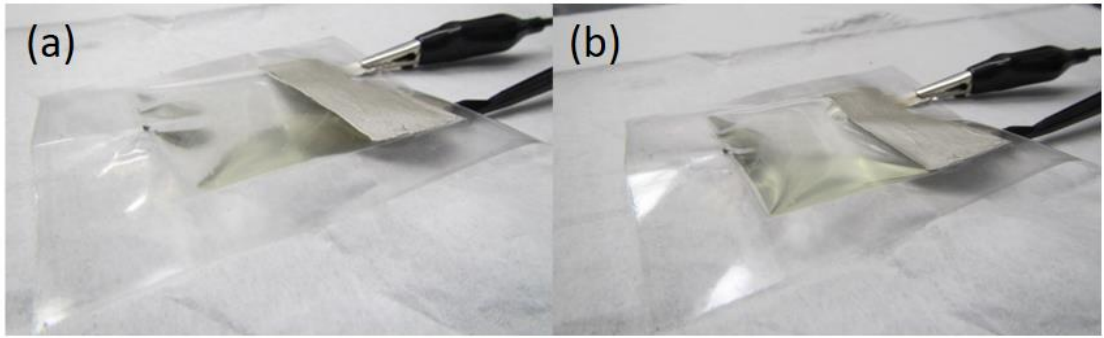

Fig. 2 — Single Element HASEL actuator fabricated using a laser engraver. Constructed from polypropylene sheets, silver paint and vegetable oil. *a. Unactuated – Voltage off. b. Actuated – Voltage applied (~10kV).*

### 1.3.1 Single Element Fabrication Procedure

Two rectangles were cut out of the clear side of the bag, making note of the side that used to be the interior (+ side from here on). The plastic was then cleaned with isopropyl alcohol to remove dust. The two layers of plastic were stacked with the + sides touching, and rubbed together to induce static. The layers were placed in the laser engraver and the parameters set according to table 1, and adjusted as necessary to get a full seal (Fig. 3).

A small port was cut open to inject vegetable oil through, and the opening was sealed, using a 3D pen set at 220°C as a heating element. The melting point of PP is around 165°C for reference. Paint both sides of the actuator, starting from the edge of the unsealed portion, with silver paint. Apply electric field between electrodes using alligator clips. This should start a zipping actuation, like that shown in Fig. 1. Around 10kV was necessary before actuation was observed.

Table 1: Laser engraver parameters used on PLS6MW (ULS).

| Operation | Power (%) | Speed (%) |
|---|---|---|
| Cut | 1 | 10 |
| Cut and Seal | 50 | 100 |

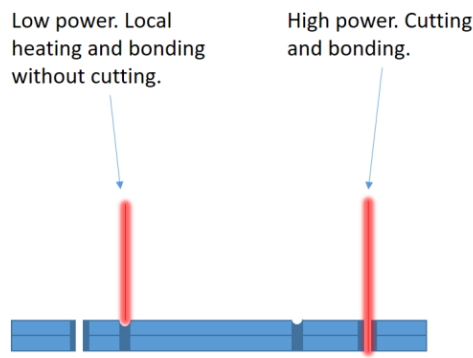

Fig. 3 — Laser engraver power and speed settings can be used to moderate local heating. This allows for controlled switching between sealing and cutting vs sealing-only operations.

The process of getting the laser parameters to the appropriate values can be delicate. Excessive heating can weaken the thermoplastic and lead to early dielectric breakdown at the edges of the fluid filled area. For this reason, the laser engraver settings are shown in Fig. 9, in the Appendix section. To repair the breakdown site, apply a heating element to seal off the damaged zone. A 3D printer pen was adjusted to 220°C to accomplish this task, and used successfully. The time required to cut and seal using the prescribed method is on the order of seconds. Using a pre-formed heating plate would require time to acquire or machine the plate, as well as cutting the sample out afterwards. Even with the 3D printer method [7], the laser beam can still move faster than a print head, reducing the time slightly on the order of seconds. It also eliminates the need to cut the sample, which reduces the chance of human error and sample variability.

A single element shell design was shown to have sealed well with this technique, even when under pressure from actuation. The fluid is pushed from the silvered area into the clear area in Fig. 4, as expected for HASEL actuation, which was illustrated in Fig. 1. The actuation force is enough to lift a small, 3D printed plastic boat (10.95 g).. Since the fluid is pushed from the silvered section into the clear section, the boat appears tilted after actuation.

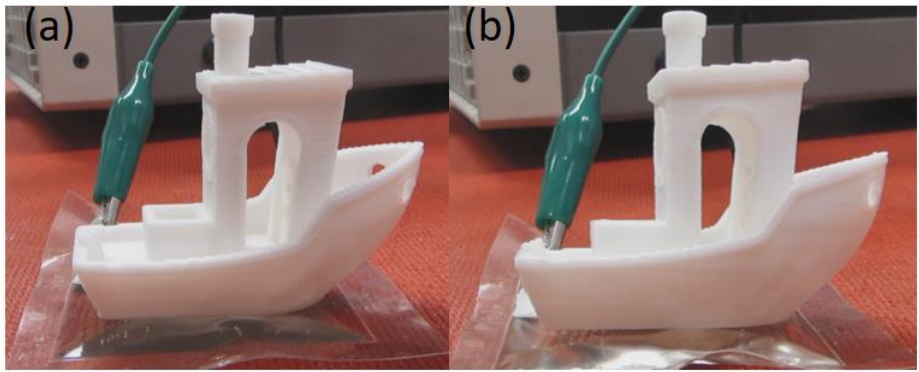

Fig. 4 — Single Element HASEL actuator lifting a 3D printed plastic boat. *a. Unactuated – Voltage off. b. Actuated – Voltage applied (~10kV)*

**HASEL LENS**

### 1.4   Optical Lens Fabrication

Building upon this technique, a HASEL actuated lens was fabricated (Fig. 5 - 8). Rather than laser cutting out the entire design, PDMS (polydimethylsiloxane) was used to form the additional chamber for the lens to the original single element and Sil-Poxy (Smooth-On) was used to connect them together. PDMS was used instead of PP since its ability to stretch allowed fluid to flow into the lens, rather than buckling which restricts flow. By applying a voltage differential across the electrodes in the pump (heat-sealed bag section) the vegetable oil is forced into the lens chamber (Fig. 5). The curvature of the fluid surface creates a visible change in the magnification of the PDMS-fluid lens.

To create the PDMS lens section, a rectangular mold (50x50x3mm) and disk (3.3mm deep, 34mm diameter) was 3D printed. The rectangular mold was filled with PDMS and the disk was placed in the center to form the cavity. Another rectangular sheet was made without a cavity. The sheets were adhered to the heat seal bag using Sil-Poxy. A similar process was used to make the lens design in Fig. 8.

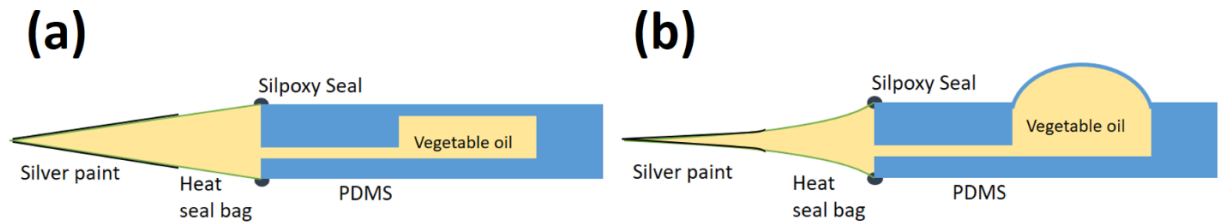

Fig. 5 — Side view schematic of the HASEL lens. The original single element functions as a pump to actuate the lens. *a. Unactuated. b. Actuated – Fluid forced from bag into lens. Lens expands out of plane.*

### 1.5 Lens Results

In the first iteration of the lens, buckling in the channel caused the fluid flow to throttle, reducing the effectiveness of the pump. This is likely due to the inextensibility of the PP bag. The buckling region is marked in Fig. 6.

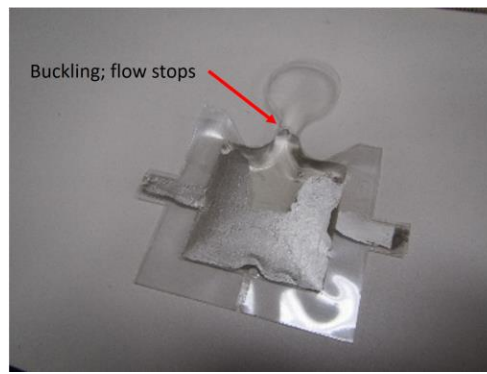

Fig. 6 — Early attempt at creating a HASEL lens, which suffered from buckling in the fluid channel. The buckling prevented the lens from actuating. Rectangular tabs are leads for applying a voltage differential across the pumping section.

By replacing the lens section with a PDMS version, a lower amount of buckling was achieved (Fig. 7). Without resolving the buckling issue, actuating the lens (and therefore magnification) would not have been possible. In Fig. 7, the magnification of letters "NRL" is shown to change by hydraulic actuation of a circular PDMS lens. This demonstrates that the fluid was able to flow into the lens section from a reservoir. Full actuation of the lens took around 8 seconds.

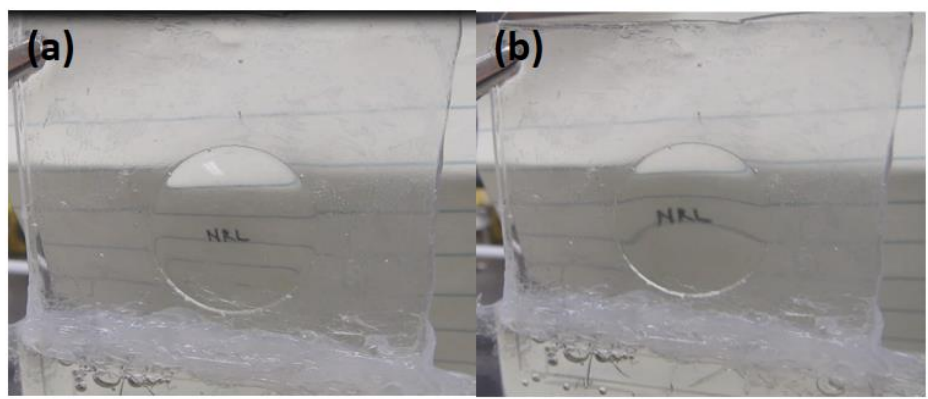

Fig. 7 — Using PDMS to form the lens section, lensing can be done by pumping from the original single element.

a. Unactuated – No magnification.  b. Actuated – Magnification of the letters "NRL".

Placing the reservoir closer to the lens in an annular configuration resulted in actuation times being brought down to around 1 second (Fig. 8).  In this design, twelve evenly spaced channels are placed radially in the PDMS that connects the fluid in the pumping section (silver ring) to the lens section in the center.  The channels in $r_1<r<r_2$ provide structural support to selectively inflate the PDMS in the lens section ($r<r_1$).  In $r_2<r<r_3$, the PP bag is painted and used to apply pressure to the fluid so that it migrates through the channels into the lens.

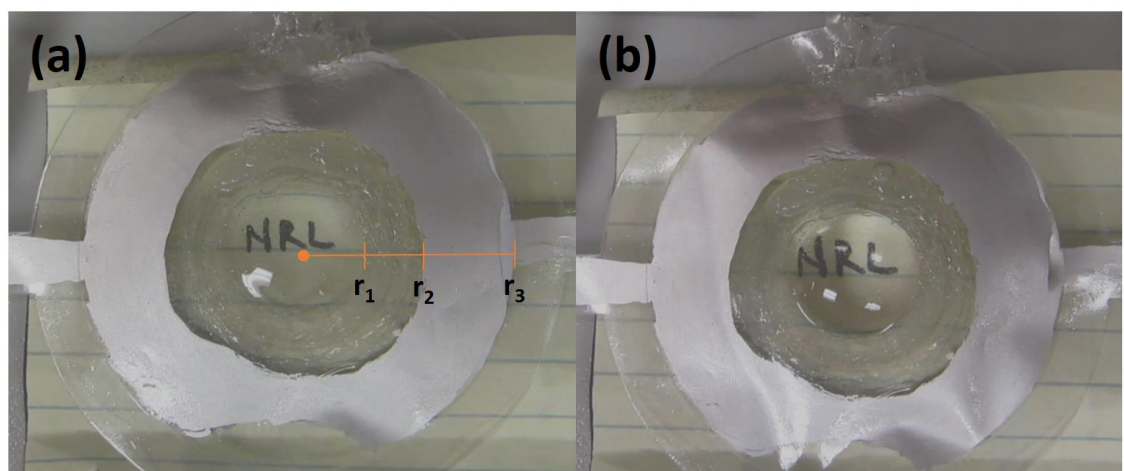

Fig. 8 — With the lens section in the center, buckling effects can be avoided.  This sample also uses a PDMS lens, which further reduced buckling.  Lens actuation was noted to be slightly faster than the design in fig 8.
a. Unactuated – No magnification.  $r<r_1$ – Lens section.  $r_1<r<r_2$ – PDMS fluid channels.  $r_2<r<r_3$ – Heat seal bag with silver electrodes.
b. Actuated – Magnification of the letters "NRL".

**CONCLUSION**

HASEL actuators allow for a compact, robust design and offer hydraulic and linear actuation. The speed, configurability and convenience of creating these actuators can be improved by using a laser engraver as a simultaneous cutting and sealing technique. As a proof of concept, a HASEL actuator was demonstrated to pump from a reservoir into a hydraulic lens. This design also incorporates soft materials only, as opposed to other designs which have rigid reservoirs. This expands upon current HASEL designs and soft lens designs, which has potential applications in compact optics as a replacement for bulky servo systems, or in biomedical devices, which often interface with other soft systems.

Future work is planned towards quantitative measurement of the focal length change of the lens and performance at different voltage levels. Additionally, improvements to the fluid used and the channel design of the lens type in Fig. 8 may lead to improved actuation times.

**APPENDIX**

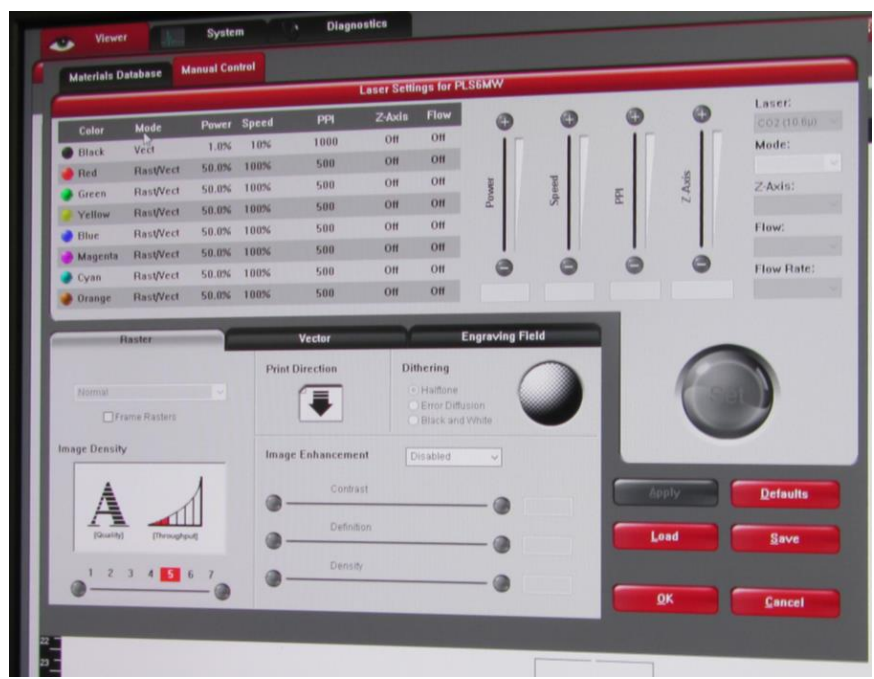

Fig. 9 — Settings in laser engraver software (UCP, ULS). Different speed and power settings are applied to lines of different colors from the imported vector drawing. The settings presented are for sealing (Black) and cutting (Red). *Shown above: Black - Vect, 1% Power, 10% Speed, 1000 PPI. Red – Rast/Vect, 50% Power, 100% Speed, 500 PPI.*